# Step-Index Optical Fibre Drawn from 3D Printed Preforms

Kevin Cook[1], Geoffrey Balle[1], John Canning[1,2,*], Loïc Chartier[1], Tristan Athanaze[1], Md Arafat Hossain[1], Chunyang Han[1], Jade-Edouard Comatti[1], Yanhua Luo[2] and Gang-Ding Peng[2]

[1] interdisciplinary Photonics Laboratories (iPL), School of Chemistry, The University of Sydney, NSW 2006, Australia;
[2] National Fibre Facility, School of Electrical Engineering and Telecommunications, University of New South Wales, Sydney, NSW 2052, Australia
*Corresponding author: john.canning@sydney.edu.au

**Optical fibre is drawn from a dual-head 3D printer fabricated preform made of two optically transparent plastics with a high index core ($NA \sim 0.25$, $V > 60$). The asymmetry observed in the fibre arises from asymmetry in the 3D printing process. The highly multi-mode optical fibre has losses measured by cut-back as low as $\alpha \sim 0.44$ dB/cm in the near IR.**

In recent work, 3D printing was shown to be an ideal technology for fabricating optical preforms from which optical fibre can be drawn [1,2]. These early fibres consisted of single material structures using air holes to guide light weakly in the core, the quality and confinement limited by the printing resolution of a low-cost desk-top 3D printer. Despite present limitations, the demonstration was clear in its potential as a technological disruptor in the manufacture of structured optical fibres – conventional methods only permit ordered periodic arrays of holes arising from the stacking process used. More arbitrary structures require considerably more challenging fabrication processes that escalate already very high costs of structured fibres. As technology improves, and new materials are integrated into the 3D printing platform at increasingly higher temperatures, this capability will extend beyond plastic into key material systems, eventually reaching silica based glass.

As important as this demonstration was, the technology used a single material system as many low cost printers involve only one material nozzle within the 3D printer. A much more significant extension of the principle will be to demonstrate dual or multi-material systems required for more conventional step-index optical fibres [3,4]. In this paper, we report the first demonstration of a step index fibre using a dual head low-cost 3D printer. The initial target is a multimode optical fibre analogous to that used in local area networks (LAN) within computer mainframe systems for example. Two materials are carefully selected to be both optically transparent whilst having different refractive indices once drawn into fibre.

The fabrication method is fused deposition modelling (FDM) where essentially a plastic filament is melted and deposited in *xy* coordinate fashion layer by layer to build up the desired structure, much like a hot glue gun put on a stage [5-7]. For this work we are using a dual head 3D printer capable of printing two materials. From previous work [1] we found that low cost printers had poor correlation between stated properties from the manufacturer and those during a printing run. Of particular note is temperature where the 3D printer-recorded temperature is measured by thermocouple near the heating source, close to but not within the nozzle area the filament traverses. This has meant that the recommended filament melting temperature for optimal printing can vary substantially from the thermocouple reading. Given that this difference can lead to differences in binding between two plastics with different melting temperatures and thermal expansion coefficients, we decided to characterize the printer head. To do this, we used an optical fibre Bragg grating ($\lambda$ =1546.7 nm, T = 35 dB) inscribed by UV within a standard silica based telecom fibre (SMF28) and stabilised by annealing ($T = 300$ $^0$C, $t = 15$ mins), to operate up to 300 $^0$C [8]. The temperature response was measured against an industrial thermocouple calibration oven. The grating was then passed through the heated nozzles of the printer and the temperature extracted along the nozzle and compared against the temperature of the printer thermocouple. The grating wavelength was tracked using a Micron Optics SM125 interrogator. Figure 1 shows the results of temperature versus position along the nozzle. The thermocouple temperatures provided by the printer are approximately 3 $^0$C lower than those measured within the nozzle. Grating interrogation provides insight into both the actual temperature and temperature distribution within the nozzle – the agreement obtained here between the measured peak temperature and that indicated by the printer thermocouple is remarkable given other, more expensive low-end printers we have measured have much larger discrepancies in excess of 10 $^0$C. More detailed studies comparing printers will be presented elsewhere.

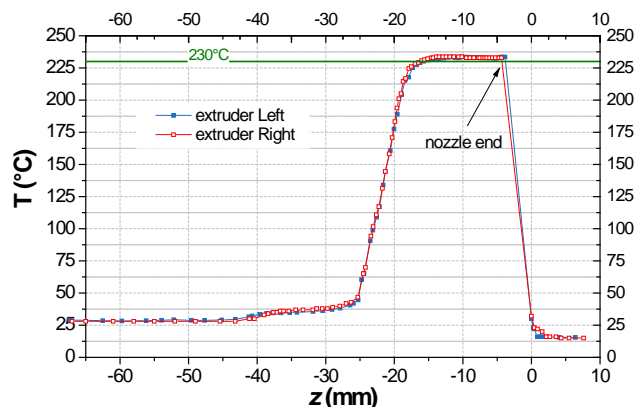

Fig. 1. Temperature, *T*, profile along the printing nozzles of the Flashforge printer as a function of position as measured by fibre Bragg grating (FBG). The nozzles are very well matched and reach a peak temperature +3 $^0$C above that measured by the extruder thermocouple (230 $^0$C).

The filaments used in this work are modified acrylonitrile butadiene styrene, ABS (or SBP [1]), for the core and modified polyethylene terephthalate glycol filament, PETG for the cladding. The UV-Vis-NIR spectra (200 to 850 nm) measured through a single filament ($\phi$ = 1.75 mm), suspended in deionized water, are shown in Figure 2. From this data, the core material can transmit much shorter wavelengths although background absorption in the green is higher. Overall transmission of both materials is excellent, where *T* > 99% through 1.75 mm towards the near

IR. The refractive index of the materials was obtained by immersing short sections in a transparent cuvette containing ortho-dichlorobenzene ($n$ = 1.569) which was diluted with various volumes of ethanol ($n$ = 1.366) to reduce the index in controlled amounts. Scattering of red HeNe laser light ($\lambda$ = 632.8 nm) through the cuvette was monitored using a power meter until it was undetectable – the mixed solution was then directly measured on a refractometer to extract the matching index of the filament. The indices of the two filaments at $\lambda$ = 633 nm were $n_{SBP}$ = (1.542 ± 0.004) and $n_{HD}$ = (1.522 ± 0.004) respectively, providing a core cladding step index of $\Delta n \sim 0.02$ in the preform prior to fibre drawing.

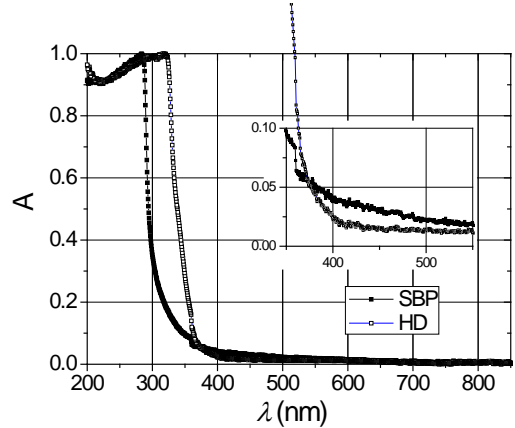

Fig. 2. UV-Vis-NIR (250 to 850 nm) transmission spectra of filaments; Inset shows the higher visible attenuation in the green of modified ABS (SBP) over the modified PETG (HD) polymer glasses.

The preform ($\phi_{core}$ = 8.00 mm, $\phi_{clad}$ = 18.60 mm, $L$ = 90 mm) was designed in AutoCAD software with two concentric cylinders suitable for 3D printing. Simplify3D was then used to format the design for 3D printing on a low-cost desktop Flashforge Dreamer printer with dual extruders. The preform was printed lengthwise along the printer bed ($T_{bed}$ = 80 $^0$C, extruder 1: $T_{SBP}$ = 230 $^0$C, extruder 2: $T_{MPETG}$ = 225 $^0$C). The printing speed was set to $v$ = 10 mm/s with an infill percentage at 100%. For this printer, such a slow speed and high infill percentage is required to maximize printing quality, minimizing air bubbles. Resolution of the printing on the $xy$-axes was set to $\Delta x = \Delta y$ = 0.4 mm and $\Delta z$ = 0.1 mm on the $z$-axis. Total printing time was $t_{print}$ = 21.5 hours. Fig. 3 (a) shows the final end-face of the printed preform, where a well-defined core is visible. Fig. 3(b) shows the end face when a $\lambda$ = 532 nm laser pointer illuminates the other end of the preform at an angle of approximately 30$^0$ to the surface normal, confirming the higher index of the modified ABS material after 3D printing.

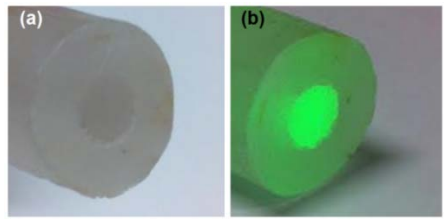

Fig. 3. Photographs of the final end-face of the printed preform: (a) without light injected and (b) with light ($\lambda$ = 532 nm, $P$ = 1 mW) injected from the other end.

The preform was drawn to fibre form using a home-made draw tower with a low temperature radiative furnace ($T_{furn}$ = 230 $^0$C) and a computer controlled feeder and spooler ($v_{feed}$ = 1.0 mm/s, $v_{spool}$ = 3500 mm/s). Approximately $L$ = 65 m of fibre was spooled. Fig. 4 (a) shows an optical microscope image of the fibre end-face cleaved with a heated razor blade. The elliptical cross-section arises from the asymmetry introduced by the lengthwise layer-by-layer printing likely exacerbated by the limitations of the home-made drawing system – an obvious way to improve the fabrication is to print vertically. A concern is the appearance of air gaps between core and cladding which will give rise to large scattering losses. This suggests that the thermal expansion coefficients (which were not known for these filaments) differed substantially and the fibre drawing process requires further optimization. The outer diameters of the fibre major and minor axes are $\phi_{OD1} \sim$ 283 µm and $\phi_{OD2} \sim$ 204 µm whilst the core diameters are $\phi_{c1} \sim$ 171 µm and $\phi_{c2} \sim$ 60 µm, respectively. With a fibre parameter $V > 60$, the fibre is inherently highly multimode. Fig. 4(b) shows a photo of the $\lambda$ = 543 nm line of a helium-neon laser launched into the fibre spool indicating overall transmission in the core is good. No differences were observed in launching two orthogonal polarized light eigenstates indicating strong mixing, unsurprising given the scattering sites at the interface and bends in the fibre.

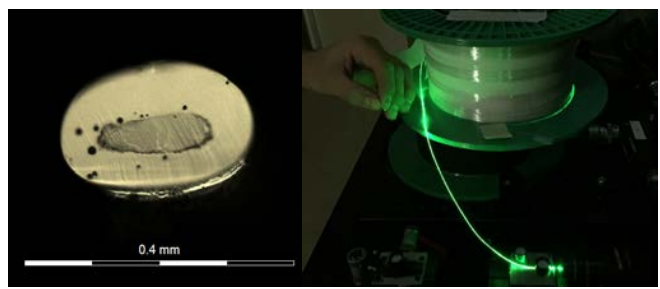

Fig. 4. Optical microscope image of the fibre end-face; (b): photo showing $\lambda$ = 543 nm light from a HeNe laser guided along the fibre.

Modal images were taken at three wavelengths by launching light into the polymer fibre ($L$ = 30 cm) from a single mode telecom fibre (SMF28) and imaging the output onto a Vidicon Vis-NIR camera. Three wavelengths were used: $\lambda$ = 543 nm (CW HeNe laser, $P_{launch}$ = 0.76 ± 0.02 mW), $\lambda$ = 1047/1052 nm (CW Nd:YLF, $P_{launch(av)}$ = 0.62 ±0.02 mW), and $\lambda$ = 1520 – 1560 nm (CW, $Er^{3+}$ doped fibre amplifier, EDFA, $P$ = 1.5 ± 0.2 mW). The fibre had a 90° bend with a radius of curvature $r \sim$ 65 mm, ensuring no forward scattering from the input end illuminated the camera. Fig. 5 (a)-(c) shows the output of the fibre for each wavelength demonstrating that light is confined to the core. Some light from the core is coupled to the cladding, due to scattering and bending at the core-cladding boundary of very high order modes. Absorption also affects the green light (Fig 2).

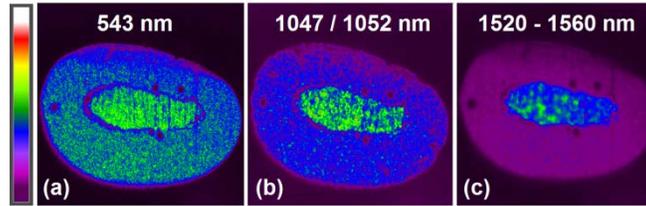

Fig. 5. Imaged fibre output for (a): $\lambda$ = 543 nm, (b): $\lambda$ = 1047 nm, and (c): $\lambda$ = (1520 – 1560) nm.

The propagation losses determined by cut-back [3] were made using the same sources. Table 1 summarises these losses – the lowest occur farthest in the near IR where $\alpha$ = 0.42 dB/cm.

**Table 1:** Losses at three wavelength by cut-back measurements.

| Source | $\lambda$ (nm) | $\alpha$ (db/cm) |
|---|---|---|
| HeNe | 543 | 0.64 ± 0.03 |
| Nd:YLF | 1047/1052 | 0.44 ± 0.02 |
| EDFA | 1520 - 2560 | 0.94 ± 0.18 |

It is interesting to note that the green light has relatively low losses despite the absorption present in the starting filament material (Fig 2), something that might suggest fibre drawing has improved through annealing and that scattering losses may be less if the band edge has moved to shorter wavelengths. Nonetheless, losses are lower at 1047 nm as expected. Losses have increased a little in the telecom window, with scattering of the air gaps being the main culprit.

In conclusion, the first step-index, core-guiding optical fibre fabricated from a 3D-printed preform is demonstrated, building on top of the recent work on structured optical fibres. Good transmission at several wavelengths including the telecom window has been demonstrated, showing the approach is suitable for production of LAN fibres for example. More work is required to reduce losses further by improving the printing process and the fibre drawing to address the core and cladding interfacial variations that seem to occur. It has been shown that highly transparent plastics available to 3D printing via fused deposition makes preform fabrication feasible on very low cost 3D desktop printers. Clearly, the results will improve dramatically on much more refined instruments and with better resolution it is not unreasonable to presuppose eventually optical fibres themselves will be made directly by 3D printing.

These 3D printing approaches also lend themselves to waveguide fabrication more broadly. The ability to print with multiple materials opens the door to tailoring the composition and refractive index of the cores and it should be straight forward to introduce all kinds of dopants into these systems and to fabricate single mode fibres. Within laboratories the capability of introducing a range of dopants into polymers and plastics that can be 3D printed also exists. For example, filaments embedded with dyes, graphene particles and other materials are already available. Given that other methods of 3D printing, including light induced polymerization and laser sintering, are in principle capable of producing preforms and even fibres, the anticipated technological disruption in optical fibre and waveguide fabrication is already here.

**Funding.** This project was primarily funded by the Australian Research Council (ARC) through DP140100975 and private funding. The National Fibre Facility is funded by ARC grants LE0883038 and LE100100098.

**Acknowledgments**. Md. Hossain acknowledges an Australian Postgraduate Award, J-E. Comatti, G. Balle, L. Chartier, and T. Athanaze acknowledge support from acknowledge support from Polytech Paris-Sud (Université Paris-Saclay) to visit *i*PL as part of their Engineering Diplomas. C. Han was supported in part by the China Scholarship Council(CSC) and the Graduate School of Xi'an Jiaotong University. Y. Luo acknowledges the support of Open Fund of State Key Laboratory for Modification of Chemical Fibers and Polymer Materials, Donghua University (LK1502). Micron Optics is acknowledged for contributing the interrogator used in this work.